\documentstyle[aps,preprint]{revtex}
\begin{document}

\title { Multi-pion correlations in high energy collisions}
\author{Q.H. Zhang}
\address{Institut f\"ur Theoretische Physik, Universit\"at Regensburg,
D-93040 Regensburg, Germany}
\vfill
\maketitle

\begin{abstract}
Any-order pion inclusive distribution for a chaotic source 
in high energy collisions are given which 
can be used in both theory and experiment to analyze 
any-order pion interferometry. 
Multi-pion correlations effects on two-pion and three-pion 
interferometry are discussed.
\end{abstract}

PACS number(s): 13.85 Hd, 05.30Jp, 12.40 Ee. 

Hanbury-Brown and Twiss\cite{HBT} were the first who applied the Bose-Einstein (BE)
correlations to measure the size of distant stars. 
The method was first applied to 
particle physics by Goldhaber et al.(GGLP)\cite{GGLP} in 1959. Since then the size 
of the interaction region has been measured by numerous experiments in high 
energy collisions using different types of particles.  
Two-pion BE correlation is widely used in high energy 
collisions to provide the information of the space-time
structure, degree of coherence and dynamics of the region where 
the pions were produced\cite{ZL}. 
 Ultrarelativistic hadronic collisions provide 
the environment for creating dozens of pions \cite{OPAL,DELPHI,NA22,UA1},  
therefore one must take into account the effects of multi-pion BE 
correlations in those processes. 
 Because the bosonic nature of the 
pion should 
affect the single and $i$-pion spectra and distort the $i (i\ge 2)$-pion 
correlation 
function,
thus, it is very interesting to analyze the 
multi-pion  BE correlation effects on $i$-pion 
interferometry\cite{LL84,Zajc87,Pratt93,CGZ95,SB,APW,CZ,CZ1}. 
On the other hand, one also want to study higher-order pion 
interferometry directly to see what additional information 
can be extracted from higher-order pion interferometry. Now this aspect 
has aroused great interests among physicists\cite{NA22,UA1,APW,CZ2,HZ97,ELB}. 
Unfortunately all present analyses of multi-pion correlations are based 
on pure ``multi-pion interferometry formula" without considering the 
higher-order pion correlation effects on the lower-order pion 
interferometry\cite{NA22,UA1,APW,CZ2,HZ97,ELB}.  Thus we urgently need 
 new multi-pion interferometry formulas which can be used 
in both theory and experiment to analyze any-order pion interferometry. 
This is the main aim of this letter. In the letter, considering 
multi-pion BE correlations we  derive 
 new multi-pion correlations formulas which can be used to analyze any-order pion 
interferometry. Those new multi-pion correlations formulas are 
structurally similar to the previous formulas but with a modified source 
functions (See Eq. (21) for details). This warrants the validity of the formulas used in 
earlier studies of higher-order pion interferometry. Although we 
only study multi-pion BE correlations 
in this Letter,   the results presented here are also held for kaon if 
the final state interactions are neglected.

The general definition of the "pure" $n$ pion correlation function 
$C_{n}({\bf p_{1}},\cdot \cdot \cdot ,{\bf p_{n}})$ is
\begin{eqnarray}
C_{n}({\bf p_{1}},\cdot \cdot \cdot, {\bf p_{n}})
=\frac{P_{n}({\bf p_{1}},\cdot \cdot \cdot, {\bf p_{n}})}
{\prod_{i=1}^n P_{1}({\bf p_{i}})}   ,
\end{eqnarray}
where $P_{n}({\bf p_{1}},\cdot \cdot \cdot, {\bf p_{n}})$ is the 
probability of observing $n$ pions with momenta $\{ {\bf p_{i}} \}$
all in the same $n$ pion event.   The n-pion momentum probability 
distribution $P_{n}({\bf p_1},\cdot\cdot\cdot, {\bf p_n})$ 
 can be expressed as\cite{Zajc87,CGZ95}
\begin{equation}
P_{n}({\bf p_1},\cdot\cdot\cdot,{\bf p_n})
=\sum_{\sigma} \rho_{1,\sigma(1)}\rho_{2,\sigma(2)}...\rho_{n,\sigma(n)},
\end{equation}
with
\begin{equation}
\rho_{i,j}=\rho({\bf p_{i}},{\bf p_{j}})=
\int d^{4}x  g_w(x, \frac{(p_{i}+p_{j})}{2}) 
e^{i(p_{i}-p_{j})\cdot x}.
\end{equation}
Here $\sigma(i)$ denotes the $i$th element of a permutation of the sequence
${1,2,3,\cdot \cdot \cdot, n}$, and the sum over $\sigma$ denotes the 
sum over all $n!$ permutations of this sequence. $g_w(Y,k)$ can be 
explained as the probability of finding a pion at point $Y$ with momentum 
$k$ which is defined as\cite{CGZ95}
\begin{equation}
g_{w}(Y,k)=\int d^{4}y  j^{*}(Y+y/2)j(Y-y/2) \exp(-ik\cdot y) ,
\end{equation}
with 
\begin{equation}
\int g_w(x,k)d^4x d\vec{k} =n_0   .
\end{equation}
Where $j(x)$ is the 
current of the pion, $n_0$ is the mean pion multiplicity without 
BE correlation\cite{Pratt93,CGZ95}.
 From Eq. (1) and Eq. (2), the pure $n$-pion 
 correlation functions can be expressed as\cite{CZ2}
\begin{equation}
C_n({\bf p_1},\cdot\cdot\cdot,{\bf p_n})=\sum_{\sigma}\prod_{j=1}^{n}
\frac{\rho_{j,\sigma{j}}}{\rho_{j,j}}.
\end{equation}
The above correlation functions are widely 
used in both experiment and theory to analyze 
multi-pion interferometry. But in high energy experiment, the pion 
multiplicity is so large that we must take into account 
the multi-pion correlations effects on lower-order pion interferometry.

For $n \pi$ events, considering the $n$ pion 
correlations effect, the $i$-pion correlation function can be defined as
\cite{CGZ95}
\begin{equation}
C_{i}^{n}({\bf p_1},\cdot \cdot \cdot,{\bf p_{i}})=
\frac{P_{i}^{n}({\bf p_{1}},\cdot \cdot \cdot, {\bf p_{i}})}{
\prod_{j=1}^{i}P_{1}^{n}({\bf p_{j}})} ,
\end{equation}
where $P_{i}^{n}({\bf p_{1}},\cdot \cdot \cdot,{\bf p_{i}})$ is 
the the normalized modified $i$-pion inclusive
distribution in $n$ pion events which can be expressed as
\begin{equation}
P_{i}^{n}({\bf p_{1}},\cdot \cdot \cdot, {\bf p_i})=
\frac{\int \prod_{j=i+1}^{n} d{\bf p_{j}}
P_{n}({\bf p_{1}},\cdot \cdot \cdot,{\bf p_{n}})}{\int \prod_{j=1}^{n} d{\bf p_{j}}
P_{n}({\bf p_{1}},\cdot \cdot \cdot,{\bf p_{n}})}.
\end{equation}

Now we define the function $G_{i}({\bf p,q})$ as\cite{Pratt93,CGZ95}
\begin{equation}
G_{i}({\bf p,q})=  \int \rho({\bf p,p_{1}}) d{\bf p_{1}} \rho({\bf p_{1},p_{2}})
d{\bf p_{2}} \cdot \cdot \cdot \rho({\bf p_{i-2},p_{i-1}})d{\bf  p_{i-1}}
\rho({\bf p_{i-1},q}).
\end{equation}

From the expression of $P_{n}({\bf p_1},\cdot\cdot\cdot,{\bf p_n})$ (Eq. (2)), 
the one-pion to three-pion inclusive distribution can be expressed as\cite{CGZ95}
\begin{eqnarray}
P_{1}^{n}({\bf p})=
\frac{1}{n}\frac{1}{\omega(n)}\sum_{i=1}^{n}G_{i}({\bf p,p})\cdot \omega(n-i),
\end{eqnarray}
\begin{eqnarray}
P_{2}^{n}({\bf p_1,\vec p_2})&=&\frac{1}{n(n-1)}\frac{1}{\omega(n)}
\sum_{i=2}^{n} [\sum_{m=1}^{i-1}G_{m}({\bf p_1,p_1})\cdot G_{i-m}({\bf p_2,p_2})
\nonumber\\
&& +G_{m}({\bf p_1,p_2})\cdot G_{i-m}({\bf p_2,p_1}) ]\omega(n-i),
\end{eqnarray}
\begin{eqnarray}
P_{3}^{n}({\bf  p_1,\vec p_2,\vec p_3})&=&\frac{1}{n(n-1)(n-2)}\frac{1}{\omega(n)}
\sum_{i=3}^{n} [\sum_{m=1}^{i-2}\sum_{k=1}^{i-m-1}G_{m}({\bf p_1,p_1})
\cdot G_{k}({\bf p_2,p_2})
\cdot G_{i-m-k}({\bf p_3,p_3})
\nonumber\\
&& +G_{m}({\bf p_1,p_2})\cdot G_{k}({\bf p_2,p_1}) \cdot G_{i-m-k}({\bf p_3,p_3})
+G_{m}({\bf p_2,p_3})\cdot G_{k}({\bf p_3,p_2})
\nonumber\\
&&\cdot G_{i-m-k}({\bf p_1,p_1})+G_{m}({\bf p_3,p_1})\cdot G_{k}({\bf p_1,p_3})\cdot 
G_{i-m-k}({\bf p_2,p_2})
+
\\
&&G_{m}({\bf p_1,p_2})\cdot G_{k}({\bf p_2,p_3})\cdot G_{i-m-k}({\bf p_3,p_1}) 
 +G_{m}({\bf p_1,p_3})
\nonumber\\
&&\cdot G_{k}({\bf p_3,p_2}) \cdot G_{i-m-k}({\bf p_2,p_1})
] \omega(n-i),
\nonumber
\end{eqnarray}
with
\begin{eqnarray}
\omega(n)=\frac{1}{n!}\int \prod_{k=1}^{n} d{\bf  p_{k}} 
P_{n}({\bf p_1},\cdot\cdot\cdot,{\bf p_n})=\frac{1}{n}\sum_{i=1}^{n} C(i)\omega(n-i),
C(i) =\int G_i({\bf p},{\bf p}) d{\bf {p}}.
\end{eqnarray}
Here $\omega(n)$ is the pion multiplicity distribution probability. 

Similar expression can be given for $i (i \le n)$ pion inclusive distribution. 
From the above method the $i$-pion correlation function 
can be calculated for $n$ pion events.  Experimentally, one usually mixes 
 all events  to analyze the two-pion and higher-order pion interferometry. 
Then the $i$ pion correlation function can be expressed as\cite{MV97,referee}
\begin{equation}
C_{i}^{\phi}({\bf  p_1},\cdot\cdot\cdot,{\bf p_i})
=\frac{N_i({\bf p_1},\cdot \cdot \cdot ,{\bf p_i})}
{\prod_{j=1}^{i}N_1({\bf p_j})}.
\end{equation}
Here the $i$-pion inclusive distribution,
$N_i({\bf p_1},\cdot \cdot \cdot,{\bf p_i})$,
 can be expressed as 
\begin{equation}
N_{i}({\bf p_1},\cdot \cdot \cdot,{\bf p_i})
=\frac{\sum_{n=i}^{\infty}\omega(n)\cdot n(n-1)\cdot\cdot\cdot (n-i+1)
P_{i}^{n}({\bf p_1},\cdot \cdot \cdot,{\bf p_i})}
{\sum_{n}\omega(n)},
\end{equation}
with 
\begin{equation}
\int N_{i}({\bf p_1},\cdot\cdot\cdot,{\bf p_i})\prod_{j=1}^{i}d{\bf p_j}
=<n(n-1)\cdot\cdot\cdot (n-i+1)>~~~.
\end{equation}
Then the one-pion to three-pion inclusive distribution read:
\begin{equation}
N_1({\bf p})=\sum_{i=1}^{\infty}G_{i}({\bf p},{\bf p}),
\end{equation}
\begin{equation}
N_2({\bf p_1},{\bf p_2})=\sum_{i=1}^{\infty}G_{i}({\bf p_1},{\bf p_1})
\sum_{j=1}^{\infty}G_j({\bf p_2},{\bf p_2})
+\sum_{i=1}^{\infty}G_i({\bf p_1},{\bf p_2})\sum_{j=1}^{\infty}G_j ({\bf p_2},{\bf p_1}),
\end{equation}
\begin{eqnarray}
N_3({\bf p_1},{\bf p_2},{\bf p_3})&=&\sum_{i=1}^{\infty}G_{i}({\bf p_1},{\bf p_1})
\sum_{j=1}^{\infty}G_j({\bf p_2},{\bf p_2})\sum_{k=1}^{\infty}
G_k({\bf p_3},{\bf p_3})
+\sum_{i=1}^{\infty}G_i({\bf p_1},{\bf p_2})\sum_{j=1}^{\infty}G_j({\bf p_2},{\bf p_1})
\nonumber\\
&&\sum_{k=1}^{\infty}G_k({\bf p_3},{\bf p_3})
+\sum_{i=1}^{\infty}G_i({\bf p_3},{\bf p_1})\sum_{j=1}^{\infty}G_j({\bf p_1},{\bf p_3})
\sum_{k=1}^{\infty}G_k({\bf p_2},{\bf p_2})
+\sum_{i=1}^{\infty}G_i({\bf p_2},{\bf p_3})
\nonumber\\
&&\sum_{j=1}^{\infty} G_{j}({\bf p_3},{\bf p_2})\sum_{k=1}^{\infty}G_k({\bf p_1},{\bf p_1})
+\sum_{i=1}^{\infty}G_i({\bf p_1},{\bf p_2})\sum_{j=1}^{\infty}G_{j}({\bf p_2},{\bf p_3})
\nonumber\\
&&\sum_{k=1}^{\infty}G_k({\bf p_3},{\bf p_1})
+\sum_{i=1}^{\infty}G_i({\bf p_1},{\bf p_3})\sum_{j=1}^{\infty}G_j({\bf p_3},{\bf p_2})
\sum_{k=1}^{\infty}G_k({\bf p_2},{\bf p_1}).
\end{eqnarray}
Similar expression for $i (i > 3)$ pion inclusive distribution can be given. 
 Following Ref.\cite{CZ,CZ1,Prattx}, we define the following function 
\begin{equation}
H_{ij}=H({\bf p_i},{\bf p_j})=\sum_{k=1}^{\infty}G_k({\bf p_i},{\bf p_j}).
\end{equation}
Then the $n$ pion inclusive distribution can be 
expressed as  
\begin{equation}
N_{n}({\bf p_1},\cdot\cdot\cdot,{\bf p_n})=\sum_{\sigma}
H_{1\sigma(1)}H_{2,\sigma(2)}\cdot\cdot\cdot H_{n,\sigma(n)}.
\end{equation}
Here $\sigma(i)$ denotes the $i$th element of a permutation of the sequence
${1,2,3,\cdot \cdot \cdot, n}$, and the sum over $\sigma$ denotes the 
sum over all $n!$ permutations of this sequence. 
One of the interesting things about Eq. (21) is that 
it is very similar to Eq. (2). The only difference is that 
Eq. (2) only contains the 
the first term  of $H_{ij}$ ( $\rho({\bf p_i,p_j})=G_1({\bf p_i,p_j})$).  
With the help of Eq. (21), the general form of $i$-pion correlation 
function (Eq. (14)) can be re-expressed as\cite{CZ2}
\begin{equation}
C_{i}^{\phi}({\bf p_1},\cdot\cdot\cdot,{\bf p_i})
=\sum_{\sigma}\prod_{j=1}^{i}
\frac{H_{j,\sigma(j)}}{H_{j,j}}.
\end{equation}

In the derivation of Eq.(22), we mention nothing about the structure of the source, 
so our results are in principle independent on the detail form of the source, i. e., 
one can use any kind source function (which may contain resonance and flow ) to study 
higher-order pion interferometry. 
Assumed that $H_{ij}=|H_{ij}|exp(i\phi_{ij})$, it is clear that 
two-pion interferometry does not depend on phase $\phi_{ij}$ while 
which exists in higher order pion interferometry. So higher-order interferometry 
can be used to extract the information of phase\cite{HZ97}. If the source 
distribution function is symmetric in the 
coordinate space we have $\phi_{ij}=0$. Then multi-pion 
correlations contain the same information as two-pion interferometry does. 
In the following, we 
will use a simple model to study the multi-pion correlations effects on 
two-pion and three-pion interferometry. Similar to ref.\cite{Zajc87,Pratt93,CGZ95},
we assume the source distribution function $g(x,p)$ as
\begin{equation}
g(x,p)=n_0\cdot (\frac{1}{\pi R^2})^{3/2}
exp(-r^2/R^2)
(\frac{1}{\pi \Delta^2})^{3/2}
exp(-p^2/\Delta^2).
\end{equation}
Here $R$ and $\Delta$ are parameters which represents the radius of 
the chaotic source and the momentum range of pions respectively. Using 
Eq.(9), Eq.(14) and Eq.(15), one can calculate any-order 
pion correlation function for any kind source 
distribution\cite{ZPC}.  Cs\"org\H o and Zim\'ayi 
have found an analytically solution for above special source distribution\cite{CZ,CZ1} which 
are quoted here:
\begin{equation}
G_n({\bf p_1,p_2})=j_n\exp\{-\frac{b_n}{2}[
(\gamma_+^{n/2}{\bf p_1}-
\gamma_{-}^{n/2}{\bf p_2})^2 + 
(\gamma_+^{n/2}{\bf p_2}-
\gamma_{-}^{n/2}{\bf p_1})^2]\}
\end{equation}
\begin{equation}
j_n = n_0^n(\frac{b_n}{\pi})^{3/2}, 
b_n = \frac{1}{\Delta^2}\frac{\gamma_+ -\gamma_{-}}{\gamma_+^n-\gamma_{-}^n}
\end{equation}
with
\begin{equation}
\gamma_{\pm} =\frac{1}{4}(x \pm 1 )^2, x = R\Delta.
\end{equation}
Using this solution, one can calculat any-order pion interferometry for 
above source distribution. The analytical solution has the advantage over the 
previous method\cite{Pratt93,CGZ95} that it can be used in theory analyses.

The two-pion interferometry for different mean multiplicity $<n>$ is shown in 
Fig.1. It is 
clear that multi-pion correlation make the radius to 
become smaller.  For three-pion interferometry, we 
choose variable $Q^2=({\bf  p_1}-{\bf p_2})^2+({\bf  p_2}-{\bf p_3})^2+
({\bf p_3}-{\bf p_1})^2={\bf q_{12}}^2+{\bf q_{23}}^2 +{\bf q_{31}}^2$ and integrate the 
other eight variables, then we have three-pion correlation 
function $C_3^{\phi}(Q)$ as 
\begin{equation}
C_{3}^{\phi}(Q)=\frac{\int N_3({\bf p_1},{\bf p_2},{\bf p_3})
\delta(Q^2-{\bf q_{12}}^2-{\bf q_{23}}^2-{\bf q_{31}}^2)
d{\bf p_1} d{\bf p_2} d{\bf p_3}}
{\int N_1({\bf p_1})N_1({\bf p_2})N_1({\bf p_3})
\delta(Q^2-{\bf q_{12}}^2-{\bf q_{23}}^2-{\bf q_{31}}^2)
d{\bf p_1} d{\bf p_2} d{\bf p_3}}.
\end{equation}
Three-pion interferometry for different $\langle n\rangle$ is shown in Fig.2. It is 
clear that as $\langle n\rangle$ becomes larger, the deviations of 
three-pion correlation from 
"pure three-pion "interferometry become larger. 
 
In the following, we will discuss the effects of resonance, flow and energy 
conservation on the above calculations. The effects of resonance on pure two-pion 
interferometry is not a new subject which has been extensive studied by 
different authors\cite{CLZ96}. 
Effects of the resonance on higher-order pion interferometry based on 
pure multi-pion interferometry formulas are recently studied by Cs\"org\H o 
in Ref.\cite{CZ2}. Because Eq.(21) is similar to Eq.(2), all analyses 
in Ref.\cite{CZ2} can be applied here.  In principle, resonance should not affected 
the multi-pion correlation formulas presented here, but due to the limited momentum 
resolution of the data we will have a modified $n$-pion BE correlation 
functions as presented in Ref.\cite{CZ2}. The basic idea is that due to the 
limited resolution of the data, the contributions from the long 
lived resonance to the correlator are concentrated at lower 
relative momentum region which is 
not resoloved by two-pion interferometry. Thus the 
interference term ($G_1({\bf p_i},{\bf p_j}),i\ne j$) 
 now mainly contains the contributions from directly emitted pions and pions from 
short lived resonances. While the single particle 
spectrum ($G_1({\bf p_i},{\bf p_i})$) is not affected by 
the two-particle momentum resolution, so the intercept 
of n-pion interferometry is smaller than the ideal value $n!$. 
Based on pure two-pion interferometry formula, one 
has found that flow make the apparent radius derived from two-pion interferometry 
to become smaller\cite{Heinz}. It is expected that due to the multi-pion BE 
correlation effects, the apparent radius 
 derived from two-pion interferometry will become more smaller.
 Considering the effects of the energy 
constraint on the multi-pion interferometry, I have derived a new multi-pion 
inclusive distribution according to the method presented in Ref.\cite{Zhang1}. 
This  new formula is  similar to Eq.(21). If we integrated 
the energy from zero to infinity we will have Eq. (21) again. According to the definition 
of Eq.(14), in the calculation of Eq.(22), we actually mix all events (with 
different multiplicities and different energies). It means that in the experiment,
one has already integrated all the possible energy of the pions, so it is not 
necessary for us to consider energy constraint effects here. 

Conclusion: In this letter, we have derived the $i$ pion inclusive 
distribution and $i$ pion correlation function for a chaotic source 
which can be 
used in experiment and theory to analyze multi-pion interferometry.  
For a simple model, multi-pion correlation effects 
on two-pion and three-pion interferometry were discussed. It was 
shown that for larger mean pion multiplicity, the deviations  
of three-pion and two-pion correlation from the 
"pure" three-pion and two-pion correlation became larger. The 
 effects of resonance, flow and energy constraint on the multi-pion 
correlation formulas were discussed. 

\begin{center}
{\bf Acknowledgement}
\end{center}
The author would like to express his gratitude to Drs. U. Heinz, Y. Pang,
H. Feldmeier,J. Knoll, D. Mi\'skowiec and V. Toneev for helpful discussions.  
This work was partly supported by 
the Alexander von Humboldt foundation in Germany.

\begin{center}
{\bf Figure Captions}
\end{center}


\begin{thebibliography}{29}
\bibitem{referee}
There are two kinds of definition of $i$-pion correlation function\cite{CZ,MV97}, one is 
the definition given in Eq.(14), the another one is 
$C_2 = \frac{P_i^{nor}(p_1,\cdot\cdot\cdot,p_i)}{\prod_{j=1}^{i}P_1^{nor}(p_j)}$
with $P_i^{nor}(p_1,\cdot \cdot \cdot,p_i) =\frac{N_i(p_1,\cdot\cdot\cdot,p_i)}
{<N\cdot\cdot\cdot(N-i+1)>}$. 
In Ref.\cite{MV97}, the author discussed those two definitions for two pion 
interferometry case and 
found that the first definition was better than the second one.
\bibitem{Prattx}
This definition can be found in the preprint version of Ref.\cite{Pratt93} 
and also can be found in Ref.\cite{CZ,CZ1} where the author defined  
it as $G(i,j)$.
\bibitem{ZPC}
The detail information about the calculation can be found 
in Ref.\cite{Pratt93,CGZ95,CZ,CZ1}
which is originated from Pratt while derived in detail in Ref.\cite{CGZ95}. On the other 
hand, Zajc gave Monte-Carlo methods\cite{Zajc87} which enable us 
to study multi-pion correlation effects on 
lower-order pion interferometry. But to my best knowledge, the amount of calculational work 
for a real model will increase astronomically for all methods. So a method which enable us to 
calculate quickly multi-pion correlations is still a debate for physicist.
\end{thebibliography}
\end {document}
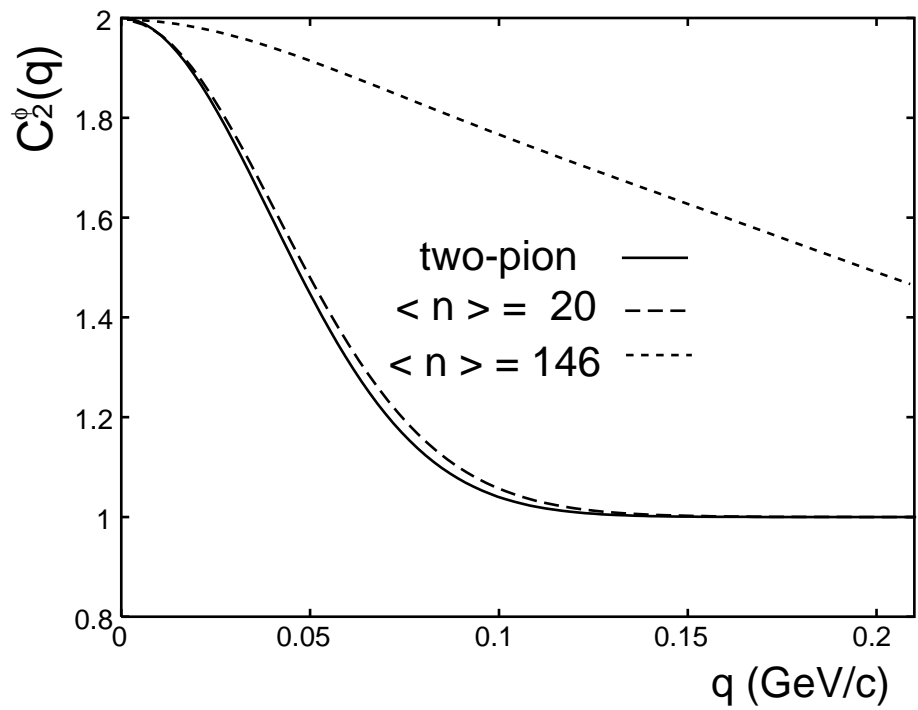
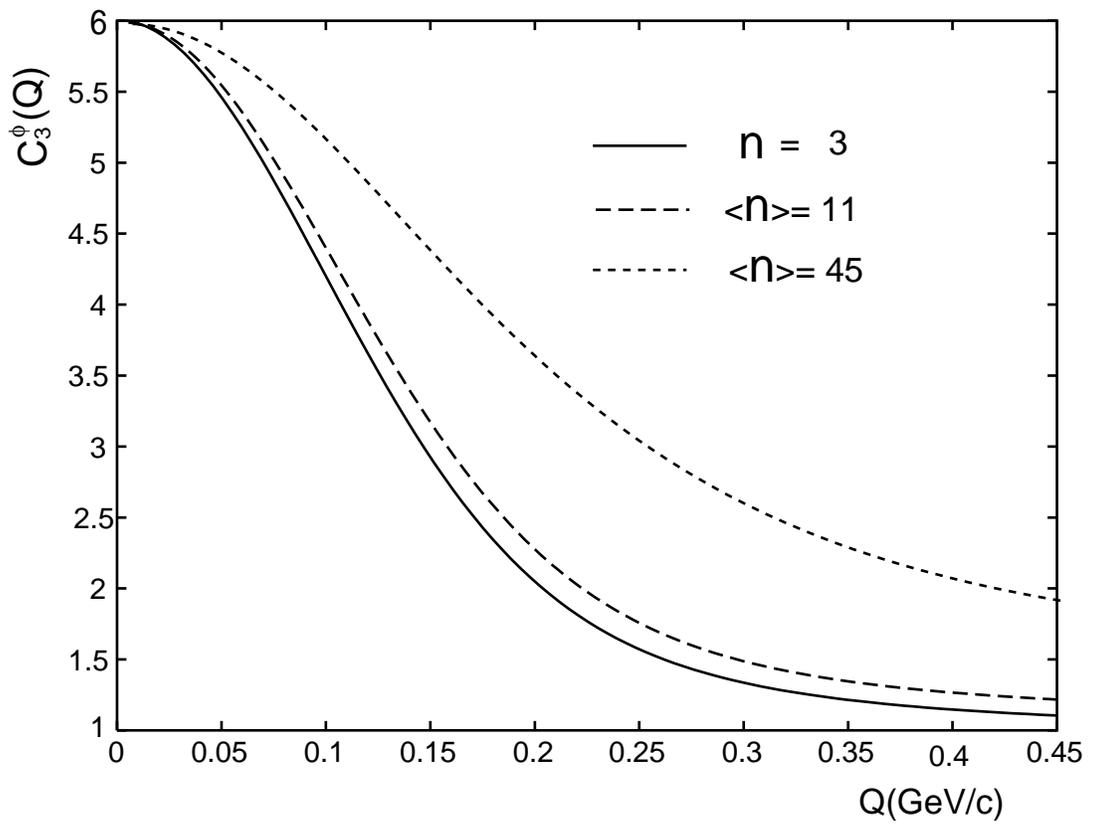
